\documentclass[useAMS,usenatbib]{mn2e}
\usepackage{graphics}  \usepackage{epsfig, color}  \usepackage{graphicx}
\usepackage{amsmath,ulem}

   \newcommand{\mbh}{M_{\rm
    bh}}  \def\ltsima{$\; \buildrel < \over \sim \;$}
\def\simlt{\lower.5ex\hbox{\ltsima}}  \def\gtsima{$\; \buildrel >
  \over\sim\;$}  \def\simgt{\lower.5ex\hbox{\gtsima}}

\def\msun{{\,{\rm M}_\odot}}  
\def\del#1{{}}

 \title[Black hole feedback in a multiphase ISM]{Black hole feedback in a
   multiphase interstellar medium}

    \author[]{Martin A. Bourne$^{1,\star}$, Sergei Nayakshin$^{1}$ and
      Alexander Hobbs$^{2}$\\  $^{1}$Department of Physics and Astronomy,
      University of Leicester, Leicester, LE1 7RH, UK\\  $^{2}$Institute for
      Astronomy, Department of Physics, ETH Zurich, Wolfgang-Pauli-Strasse 16,
      CH-8093 Zurich, Switzerland\\  $^{\star}$ {E-mail:~} {\rm
        martin.bourne@le.ac.uk} } 

\begin{document}
\date{Received} \pagerange{\pageref{firstpage}--\pageref{lastpage}}
\pubyear{2013} \maketitle
\label{firstpage}
\maketitle

\begin{abstract}
 Ultrafast outflows (UFOs) from supermassive black holes (SMBHs) are thought
 to regulate the growth of SMBHs and host galaxies, resulting in a number of
 observational correlations. We present high-resolution numerical simulations
 of the impact of a thermalized UFO on the ambient gas in the inner part of
 the host galaxy.  Our results depend strongly on whether the gas is
 homogeneous or clumpy. In the former case all of the ambient gas is driven
 outward rapidly as expected based on commonly used energy budget arguments,
 while in the latter the flows of mass and energy de-couple. Carrying most of
 the energy, the shocked UFO escapes from the bulge via paths of least
 resistance, taking with it only the low-density phase of the host. Most of
 the mass is however in the high-density phase, and is affected by the UFO
 much less strongly,  and may even continue to flow inwards. We suggest that
 the UFO energy leakage through the pores in the multiphase interstellar
 medium (ISM) may explain why observed SMBHs are so massive despite their
 overwhelmingly large energy production rates. The multiphase ISM effects
 reported here are probably under-resolved in cosmological simulations but may
 be included in prescriptions for active galactic nuclei feedback in future
 simulations and in semi-analytical models.
\end{abstract}

\begin{keywords}
galaxies: evolution, galaxies: ISM, galaxies: active, methods: numerical
\end{keywords}

\section{Introduction}
Observational correlations between the mass of supermassive black holes
(SMBHs) and their host galaxy, such as the $M-\sigma$ relation
\citep{Ferrarese00, Gebhardt00, Tremaine02} link the evolution of the SMBH and
their host bulge. Feedback \citep[e.g.,][]{SilkRees98} , in the form of ultra
fast outflows (UFOs), has been invoked to explain and derive the $M-\sigma$
relation from analytical arguments \citep{King03, king05}. The model is very
attractive due to its simplicity, reliance on common sense physics (Eddington
limit, escape velocity and radiation momentum outflow rate arguments),
observational analogy to outflows from massive stars (that are also near their
Eddington limits), and finally direct observations of UFOs in nearby bright
AGN \citep{PoundsEtal03a,KP03,TombesiEtal10,Tombesi2010ApJ,PoundsVaughan11a}.

Assuming a homogeneous gas distribution following a singular isothermal sphere
(SIS) potential (e.g., $\S 4.3.3$b in \citeauthor{BT08}, \citeyear{BT08}),
\citet{King03} shows that within the inverse Compton (IC) cooling radius,
$R_{\rm IC}\sim 500 M_{8}^{1/2}\sigma_{200}$ kpc (where $M_8$ is the SMBH mass
in units of $10^{8}\msun$ and $\sigma_{200}$ is the velocity dispersion in the
host, $\sigma$ in units of $200$ km s$^{-1}$; \citet{ZK12b}), the wind shock,
which develops when the UFO collides with the interstellar medium (ISM), can
cool effectively via IC scattering.  Most of the thermalized wind kinetic
energy is lost to this radiation, and only the pre-shock ram pressure impacts
the ISM. By considering the equation of motion of the swept up ISM shell,
\citet{King03} derived the mass that the SMBH had to attain in order to clear
the host galaxy's gas. Beyond the cooling radius, $R_{\rm IC}$, the wind shock
cannot cool effectively and retains the wind kinetic energy in the form of
thermal energy and the outflow becomes energy driven. This regime is much more
effective at clearing a galaxy of gas.  

The model of \citet{King03} assumes the electrons and ions in the shock share
a single temperature at all times, initially the shock temperature $T_{\rm
  sh}\sim 10^{10}$K. However, \cite{FQ12a} have shown that, due to the high
temperature and low density of the shocked wind, the electron-ion energy
equilibration time-scale is long compared with the Compton time-scale.  This
would imply that the electron temperature is much lower than the ion
temperature, i.e. $T_{\rm e}\ll T_{\rm ion}$. \citet{Bourne13}  point out an
observational test to distinguish between outflows with a one-temperature
($1T$; $T_{\rm e} = T_{\rm i}$) or two-temperature ($2T$) structure, and
conclude preliminarily that X-ray observations broadly support the findings of
\citet{FQ12a}.  This would however lead to significant implications for AGN
feedback on host galaxies:  most of the UFO's kinetic energy, carried by the
ions, is then conserved rather than radiated away. The cooling radius, $R_{\rm
  IC}$, becomes negligibly small on the scale of the host galaxy, and the
outflow is essentially always in the energy conserving phase. Based on
spherically symmetric analytical models \citep[e.g.,][]{king05}, even black
holes $\sim 100$ times below $M_{\sigma}$ could clear a galaxy of its gas. It
is then not clear (i) how black holes manage to grow so massive, and (ii) why
momentum-conserving flows provide such a tight fit to the observed $M-\sigma$
relations \citep{King03}.

Several recent additional numerical and analytical results however call the
spherically symmetric models of AGN feedback into question. In the context of
the physically related problem of stellar feedback, \cite{H-CMurray09}
modelled the structure of a hot bubble inflated by a cluster of young stars in
Carina Nebula, and have shown that the models assuming spherical symmetry do
not explain the observational data. At the same time, a model in which the
ambient ISM is clumpy accounts for observations much
better. \cite{H-CMurray09} build a toy analytical model in which a significant
fraction of the energy inside of the hot bubble is lost via advection, e.g.,
adiabatic expansion energy losses, rather than radiative energy losses (which
can be directly observed in X-rays in the case of Carina Nebula, and are much
lower than expected in the spherically symmetric models). Physically, the
authors argue that the compressed shell of a multiphase ISM has pores through
which the hot gas escapes. This deflates the bubble and allows a much better
explanation of the bubble size, age and luminosity.

\cite{RoggersPittard13} have recently performed 3D numerical simulations of a
supernova exploding inside an inhomogeneous giant molecular cloud, and found
results consistent with that of \cite{H-CMurray09}: the densest molecular
regions turned out to be surprisingly resistant to ablation by the hot gas
which was mainly escaping from the region via low density channels.

For the AGN feedback problem that we study here, \citet{wagner12} have
found very similar results when studying the interaction of an AGN jet with
the multiphase ISM. Furthermore, \citet{Wagner13} studied the interaction of a
wide-angle outflow with an inhomogeneous ambient medium, finding again that
hot gas mainly streams away through channels between the cold clouds; the
latter are impacted by the momentum of the UFO only. These authors also
concluded that the opening angle of the UFO at launch appears secondary, since
interactions of the UFO with the intervening clouds isotropize the hot bubble,
so that result of a jet and an UFO running into the inhomogeneous ISM may
actually be much more similar than often assumed.

In an analytical study, \citeauthor{Nayakshin14} (\citeyear{Nayakshin14}, hereafter N14) also argued that
most of the UFO energy leaks out of the porous bulge via the low-density
voids, and that the cold gas is affected only by the ram pressure. He argued
that the densest cold clouds may continue to feed the AGN via the `chaotic
accretion mode' \citep{HobbsEtal11} despite the AGN blowing an energy-driven
bubble into the host galaxy, and that the balance between the ram pressure of
the UFO on the clouds and cloud self-gravity leads to an $M-\sigma$
correlation very similar in functional form to that of \cite{King03}. 

Furthermore, \citeauthor{Zubovas14} (\citeyear{Zubovas14}, ZN14 hereafter) presented numerical simulations
of AGN feedback impacting elliptical, initially homogeneous ambient gas
distributions and showed that the UFO energy escapes via directions of least
resistance (along the minor axis of the ellipsoid). They additionally
presented a toy analytical model, similar in spirit to that of
\cite{H-CMurray09}, which showed that the SMBH growth stops when the SMBH
reaches a mass of the order of the \cite{King03} result. In this paper we
investigate these ideas further numerically. We set up a hot bubble of shocked
UFO gas bounded by either one- or two-phase ambient gas, and then study the
resulting interaction. Our multiphase gas is produced by evolving a Gaussian
random velocity field as is frequently done in numerical models of star
formation inside turbulent molecular clouds \citep{Bate09}, similar to earlier
work by \cite{HobbsEtal11}.  

 Our numerical methods and initial conditions differ substantially from that
 of \cite{wagner12,Wagner13} and ZN14, but results are qualitatively
 similar. We also find that most of the UFO energy is carried away by hot low
 density gas escaping the innermost regions of the host via paths of least
 resistance, which exists in the clumpy ISM in abundance
 \citep[e.g.,][]{McKeeOstriker77}.  Most of the gaseous mass in our models is
 in the high-density cold phase of the ISM that occupies a small fraction of
 the host's volume, and for this reason our host galaxies turn out to be much
 less vulnerable to AGN feedback than could be thought based on the energy
 budget arguments alone.

\section{Simulation Set-up}
\subsection{Numerical method}
The simulations presented here make use of a modified version of the
N-body/hydrodynamical code GADGET-3, an updated version of the code presented
in \citet{Springel05}. We implement the SPHS\footnote{Smooth particle
  hydrodynamics with a high-order dissipations switch.} formalism as described
in \citet{Read10} and \citet{ReadHayfield12}, in order to correctly treat
mixing within multiphase gas, together with a second-order Wendland kernel
\citep{Wendland95,DehnenAly12} with 100 neighbours.  The SPHS algorithm was
developed for the express purpose of capturing instabilities such as
Kelvin-Helmholz and Rayleigh-Taylor, and has been demonstrated as robust in
many test problems \citep{ReadHayfield12} and full galaxy formation
simulations \citep{HobbsEtAl13}. The simulations are run in a static
isothermal potential with the total mass of the potential within radius R
following:
\begin{equation}
  M_{\rm pot}(R)= \frac{M_{\rm a}}{a} {R}\;,
\label{MRDM}
\end{equation}
where $M_{\rm a}=5\times10^{10}\msun$ and $a=4$kpc. The potential is softened
at small radii in order to avoid divergence in the gravitational force as $R$
tends to zero. The one dimensional velocity dispersion of the potential is
$\sigma_{\rm pot} = (GM_{\rm a}/2a)^{1/2}\simeq 164$ km s$^{-1}$. In all
simulations we use an ideal equation of state for the gas, the gas pressure is
given by $P = \rho k_{\rm B}T/\mu m_{\rm p}$, where $\rho$ is the gas density,
$k_{\rm B}$ is the Boltzmann constant, $T$ is the gas temperature and
$\mu=0.63$ is the mean molecular weight. An optically thin radiative cooling
function for gas ionized and heated by a quasar radiation field (assuming a
fixed black hole luminosity of $L_{Edd}=2.5\times 10^{46}$ erg s$^{-1}$) as
calculated by \citet{sazonovetal05} is used for $T>10^{4}$ K. Below $10^{4}$
K, cooling is modelled as in \citet{Mashchenko08}, proceeding through fine
structure and metastable lines of C, N, O, Fe, S and Si. For simplicity, we
fix metal abundances  at solar metallicity.  We impose a temperature floor of
$100$ K.

 Gas particles are converted into star particles according to a Jeans
 instability condition. SPH particles with density above a critical density of
\begin{equation}
\rho_{crit} = \rho_{thresh} + \rho_{J}
\end{equation}
are turned into star particles, where $\rho_{thresh}=10^{-20}$ g cm$^{-3}$ and
$\rho_{J}$ is the local Jeans density given by,
\begin{equation}
\rho_{J} = \left(\frac{\pi k_{B}T}{\mu
  m_{p}G}\right)^{3}\left(n_{ngb}m_{sph}\right)^{-2}\simeq 1.17\times
10^{-18}T_{4}^{3} \text{g cm}^{-3}
\label{sfr}
\end{equation}
where $T_{4}=T/10^{4}$ K, $n_{ngb}=100$ is the typical number of neighbours of
an sph particle and $m_{sph}$ is the SPH particle mass. The $\rho_{thresh}$
term ensures that only high-density gas is converted into star particles
whilst the second term is the local Jeans density and ensures that stars only
form in gas that is unstable towards gravitational collapse\footnote{Strictly
  speaking in order to properly follow the collapse of gas one should be able
  to resolve the local Jeans mass, $M_{J}$, i.e. $n_{ngb}m_{sph} < M_{J}$
  \citep{Whitworth98}. Gas with $T=T_{floor}=100$ K has $\rho_{J}\simeq
  10^{-24}$ g cm$^{-3}$ leading to some gas having $\rho > \rho_{J}$ but not
  being converted into stars and hence we are not resolving the Jeans mass of
  this gas. However for the purpose of these simulations we are not particular
  interested in studying star formation in detail and the number of particles
  for which the above condition is true is negligibly small.}. Removing
high-density gas aids in reducing the computation time by removing particles
that would otherwise have prohibitively short time-steps. Each newly formed
star particle has the same mass as the original gas particle and only interact
with other particles through gravity. 

\begin{figure}
\psfig{file=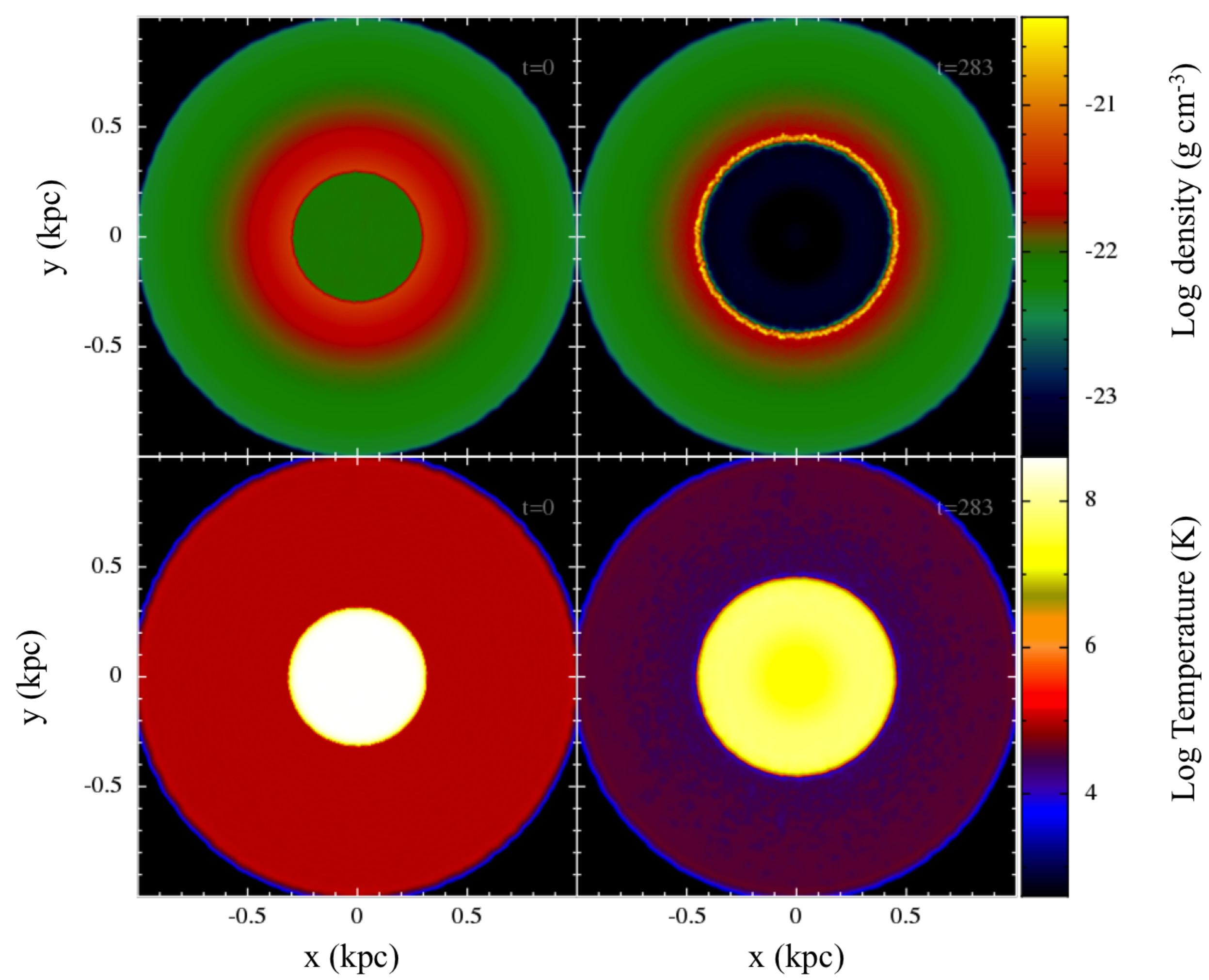,width=0.5\textwidth,angle=0}
  \caption{Density (top panel) and temperature (bottom) slices through z=0
    plane at time t=0 and $t\simeq 282$ kyr for homogeneous initial condition
    simulation H1.}
	\label{time-evo-uniform}
\end{figure}

\begin{figure*}
\psfig{file=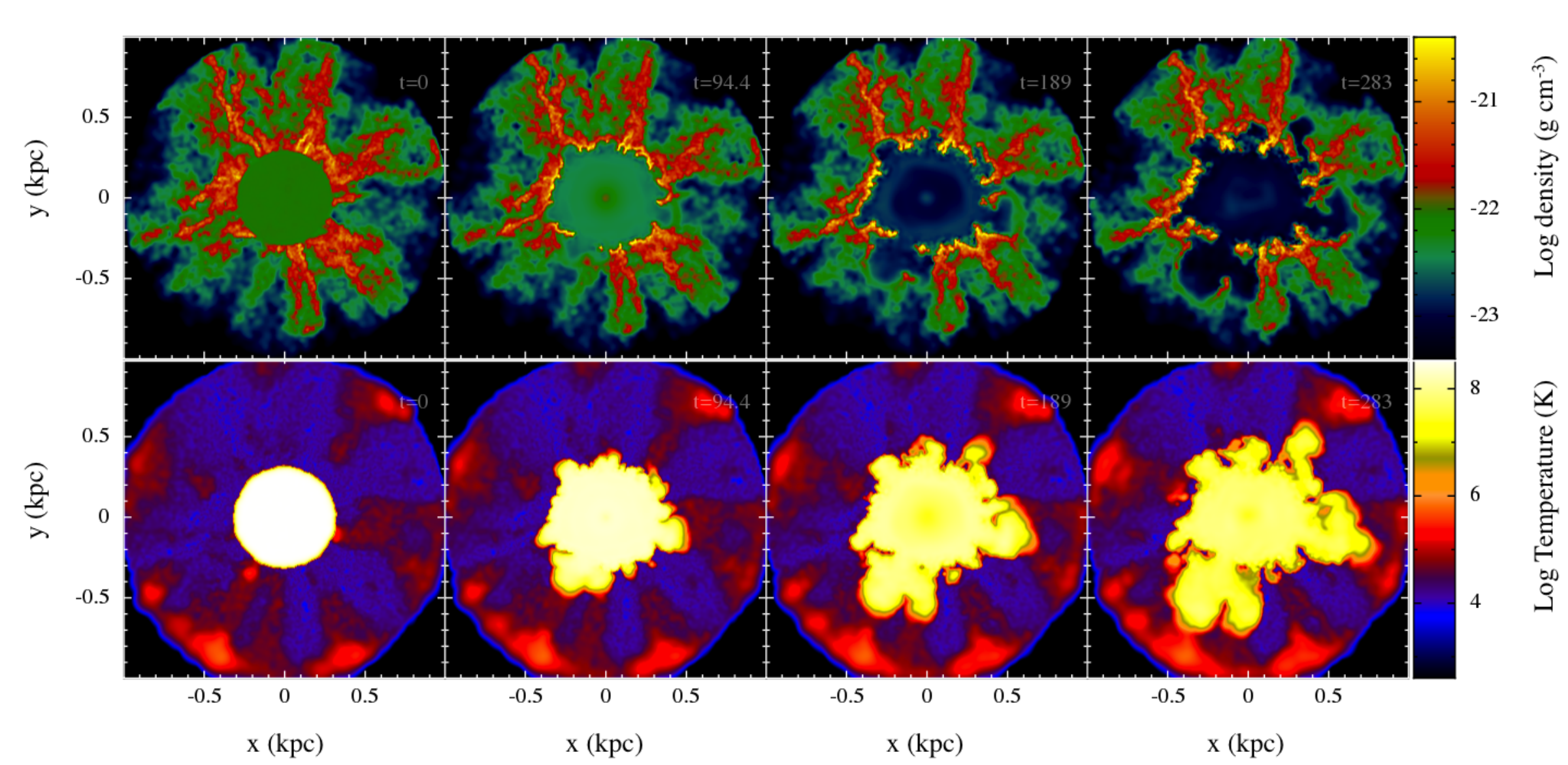,width=1\textwidth,angle=0}
  \caption{Same as Fig. \ref{time-evo-uniform} but for the turbulent initial
    condition simulation T1. Density (top panel) and temperature (bottom
    panel) slices through z=0 plane evolving in time from t=0 to $t\simeq 282$
    kyr in steps of $\sim 94$ kyr, from left to right, respectively.}
	\label{time-evo-rho}
\end{figure*}

\subsection{Initial conditions}

 Simulation of isolated galaxies by definition does not model gas inflows into
 galaxies from larger scales, and therefore idealized initial conditions for
 the ISM of the host must be used. There is a considerable freedom in choosing
 these initial conditions. In W12 and W13, cold, high-density clumps in
 hydrostatic equilibrium with the hot, low-density phase are introduced at the
 beginning of the simulation. The initial velocity of the gas is zero
 everywhere.

 In the current paper, however, since the epoch we are interested is one of a
 rapid SMBH growth and star formation in the host galaxy, the ambient gas may
 be in a very dynamical non-equilibrium state, which we model with an imposed
 turbulent velocity flow. In doing so we are inspired by numerical studies of
 star formation in molecular clouds \citep[e.g.,][]{BateEtal03}. In practice,
 our method for generating two-phase initial conditions is based on earlier
 work by \citet{HobbsEtal11}, where the importance of high-density gas clumps
 for SMBH {\it feeding rather than feedback} was studied. 

We seed a sphere of gas (cut from a relaxed, glass-like configuration) with a
turbulent velocity field following \citet{DubinskiEtAl95}. A Kolmogorov power
spectrum is assumed, $P_{\rm v}(k)\sim k^{-11/3}$, where $k$ is the
wavenumber. Gas velocity $\vec{v}$ can be defined in terms of the vector
potential $\vec{A}$, whose realization is also a power-law with the cutoff at
$k_{\rm min}$. Physically the small scale cut-off $k_{\rm min}$ defines the
largest scale, $\lambda_{\rm max}=2\pi /k_{\rm min}$, on which turbulence is
likely to be driven.  Here we set $k_{\rm \rm min} \simeq 1/R_{\rm out}$, as
the shell becomes distorted for larger $\lambda_{\rm max}$. The statistical
realization of the velocity field is generated by sampling the vector
potential $\vec{A}$ in Fourier space, drawing the amplitudes of the components
of $\vec{A_{\rm k}}$ at each point $\left(k_x, k_y, k_z\right)$ from a
Rayleigh distribution with a variance given by $<|\vec{A_{\rm k}}|^{2}>$ and
assigning phase angles that are uniformly distributed between $0$ and
$2\pi$. Finally, we take the Fourier transform of $\vec{v_{\rm
    k}}=i\vec{k}\times\vec{A_{\rm k}}$ to obtain the velocity field in real
space.  

The gas initially follows the SIS potential (meaning that $\rho(R) \propto
R^{-2}$) from $R_{\rm in}=0.1$ kpc to $R_{\rm out}=1$ kpc with a gas mass
fraction $f_{\rm g}=M_{\rm g}/(M_{\rm g}+M_{\rm pot})=0.5$, where $M_{\rm g}$
and $M_{\rm pot}$ are the gas and potential mass within the shell $0.1\leq
R\leq 1$ kpc, respectively. In order to avoid particles at small radii with
prohibitively small time steps we add a sink particle at the centre of the
simulation domain with $M_{\rm sink}=2\times 10^{8}\msun$ ( $\sim
M_{\sigma}/2$). The turbulent velocity is normalized such that the
root-mean-square velocity, $v_{\rm turb}\simeq \sigma \simeq 232$ km s$^{-1}$,
where $\sigma\simeq (GM_{\rm a}/2a(1-f_{\rm g}))^{1/2}$ is the velocity
dispersion of the potential plus gas component. 

The initial gas temperature is set to $T\simeq 1\times 10^{6}$ K, such that
the shell is marginally virialized, i.e; $(E_{\rm turb}+E_{\rm therm})/|E_{\rm
  grav}|\sim 1/2$, where $E_{\rm turb}$ and $E_{\rm therm}$ are the total
turbulent kinetic energy and total thermal energy of the gas respectively and
$E_{\rm grav}$ is the gravitational potential energy of the system.  

The system is allowed to evolve under the action of the turbulent velocity
field for time $\sim\tau_{\rm dyn}/3 = R_{\rm out}/3\sigma$, allowing the
density inhomogeneities to grow. The resulting gas shell is then re-cut to
have an inner radius $R_{\rm in}=0.3$ kpc and outer radius $R_{\rm out}=1$
kpc. The total gas mass is $M_{\rm g}\simeq 5.9\times 10^{9}\msun$,
corresponding to a gas fraction of $f_{\rm g}\simeq 0.4$ and giving a velocity
dispersion for the system (gas + potential within the shell) of $\sigma\simeq
212$ kms$^{-1}$.  The total number of particles in the gas shell is $N_{\rm
  gas}\simeq 2.6\times 10^{6}$ with a particle mass $m_{\rm gas}\simeq
2250\msun$. 

Typical parameters for an UFO give a velocity $v_{\rm out}\sim 0.1$ c, mass
outflow rate $\dot{M}_{\rm out}\sim 0.1\msun$ yr$^{-1}$ and kinetic energy
flux $\dot{M}_{\rm out}v^{2}/2\simeq 0.05L_{\rm Edd}$.  Modelling a continuous
ejection of fast wind particles by SPH is not currently feasible: at our
present mass resolution (which is much higher than a typical cosmological
simulation), a single SPH particle accounts for all of the UFO mass over $\sim
22.5$ kyr. Fortunately, it is the total energy budget of the hot shocked wind
bubble and not its minuscule mass that determines the strength of the bubble's
impact on the ambient medium \citep[the mass of the UFO is so small compared
  to the host galaxy that it does not even enter in the analytic
  theory;][]{King10b}. Therefore we rescale the properties of the UFO
particles, keeping the hot bubble's energy at a desirable value but increasing
the outflow's mass, to be able to model the thermalized UFO hydrodynamically
and with a reasonable numerical resolution. In particular, the UFO thermalized
in the reverse shock is introduced in the initial condition as a hot spherical
bubble of radius $R_{\rm bub}=0.3$ kpc centred on the sink particle. We have
tested different bubble masses and find that, qualitatively, the main
conclusions of our paper remain unchanged.

The initial gas density and temperature are assumed constant throughout the
bubble, as expected \citep{FQ12a}. The temperature and mass of the bubble are
determined based upon the desired energy ratio between the hot bubble and the
ambient gas component:
\begin{equation}
E_{\rm r}=\frac{E_{\rm H}}{E_{\rm a}}=\frac{M_{\rm H}c_{\rm s}^{2}}{M_{\rm
    a}\sigma^{2}}
\end{equation}
where $E_{\rm H}$ and $E_{\rm a}$ are the energy in the hot bubble and the
ambient gas, respectively, $M_{\rm H}$ and $M_{\rm a}$ are the total mass in
the hot and cold component, respectively, $c_{\rm s}$ is the sound speed in
the hot bubble and $\sigma\simeq 212$km s$^{-1}$ is the velocity
dispersion. All simulations presented in this paper use $c_{\rm s}\simeq 3000$
km s$^{-1}$ and $E_{\rm r}=5$; the main conclusions of our paper are
independent of $E_{\rm r}$ as long as $E_{\rm r}\gg 1$, as expected for
AGN-inflated feedback bubbles \citep{King10b}. The left-most panels in
Fig. \ref{time-evo-rho} show the initial density and temperature structure of
the system.  

As well as the runs with a turbulent medium, we have a control simulation that
has not been seeded with turbulence to contrast the outcomes. The radial gas
distribution of the control run follows the same profile as the turbulent
shell {\it before} relaxation, so that the gas is homogeneous, but has a mass
equal to that of the turbulent shell {\it after} relaxation. The initial
radially binned gas distribution is hence identical for the homogeneous and
turbulent runs save for a slight evolution during relaxation of the latter
runs as described above (compare the dashed red and blue curves in
Fig. \ref{rad-dist}). 

 It should also be noted that the control run has a low initial temperature
 $T\simeq 10^{5}$ K, which is subvirial in order to ensure that the gas
 remains homogeneous during the simulation, which is the regime we wish to
 study here. Further, since there is no imposed turbulent velocity field that
 would develop into the turbulent multiphase ISM, there is no need to relax
 this initial condition before applying the hot bubble. For this reason, the
 gas has a zero initial velocity in the homogeneous control run, unlike the
 turbulent run. This difference in initial conditions has a very minor effect
 on the final outcome of the simulations because the radial velocity gained by
 the gas in the homogeneous run is much larger than the difference in the
 initial velocities in the two runs.

 In what follows we refer to the simulations as the turbulent (T1) and control
 (homogeneous, H1) runs, respectively. In order to study the direct impact of
 the hot bubble on the ambient gas and/or to avoid confusion due to the dense
 gas phase shielding lower density gas behind it (at larger radii), a number
 of figures only include the SPH particles that were within $0.3\leq R\leq
 0.35$ kpc at $t=0$ kyr. Behaviour of gas initially at larger radii will
 nevertheless be discussed in some of the figures below.

\section{Feedback on turbulent versus homogeneous medium}

\begin{figure}
\psfig{file=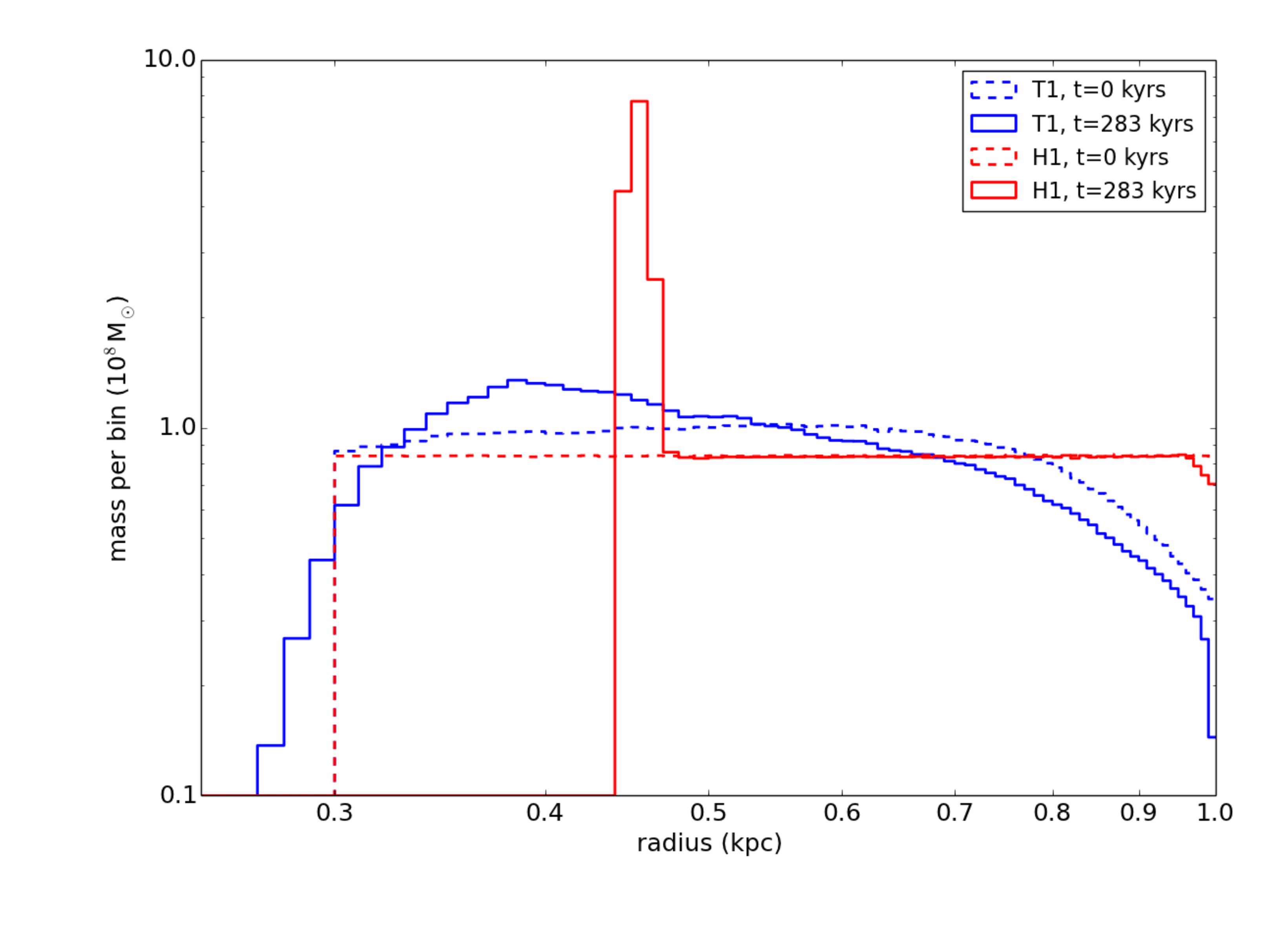,width=0.5\textwidth}
	\caption{Histogram of the gas mass in radial bins. The blue and red
          lines are for the turbulent clumpy (T1) and homogeneous (H1) gas
          distributions, respectively. The dashed and solid lines correspond
          to times $t=0$ kyr and $t\simeq 283$ kyr, respectively. Note how
          little the clumpy distribution evolves: if anything, gas continues
          to accumulate in the innermost region, whereas it is completely
          blown away in the H1 run.}
	\label{rad-dist}
\end{figure}

\begin{figure}
\psfig{file=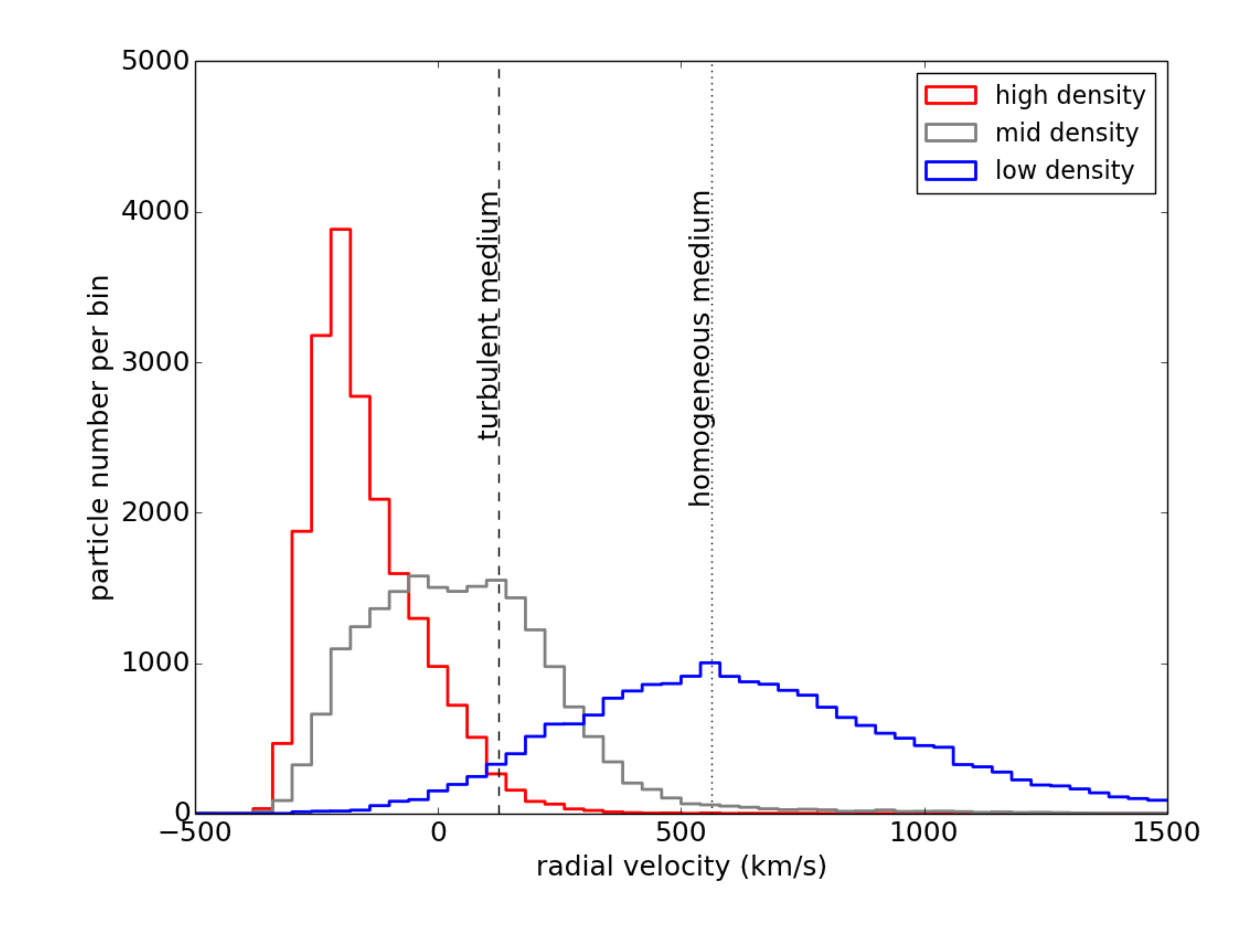,width=0.5\textwidth}
	\caption{Histogram of the radial velocity distributions $t=70.8$ kyr
          for SPH particles that belong to one of the three representative
          density groups, i.e., the highest $10\%$, around the logarithmic
          mean and the lowest $10\%$ of SPH particle densities, as labelled in
          the inset. Particles selected were within $R\leq 0.35$ kpc at $t=0$
          kyr, as explained in the text}
	\label{vel-dist}
\end{figure}

  Fig. \ref{time-evo-uniform} shows density (top) and temperature (bottom)
  slices at time $t=0$ (left) and $t\simeq 283$ kyr (right) for the
  homogeneous density run, H1. Fig. \ref{time-evo-rho} shows the same quantities
  at four different times for the turbulent initial condition simulation
  T1. The times of the first and the last snapshots are the same as those for
  Fig.  \ref{time-evo-uniform}.

It is immediately obvious that the homogeneous ambient density case, H1,
produces a ``boring'' spherically symmetric, dense, shell that is expanding
under the pressure of the hot bubble in the middle. The bubble also remains
spherically symmetric.\footnote{There may be small scale \cite{Vishniac1983}
  instabilities developing on the surface of the bubble \citep{NZ12}, but
  these instabilities grow slower than the shell is driven outward in this
  energy-conserving situation.} Importantly, the bubble drives {\it all} of
the ambient gas encountered outward at a high velocity.

 This is in stark contrast to the turbulent run  as can be seen in Fig.
 \ref{time-evo-rho}. The expansion of the hot bubble into the ambient phase
 occurs along the paths of least resistance. The low-density ambient phase is
 swept up and pushed out, while the high-density gas suffers  a much smaller
 positive radial acceleration and little (if any) gain in temperature. Some
 compression and ablation of the cold dense medium does occur, but most of it
 survives the bubble's passage intact. 

 Fig. \ref{rad-dist} highlights the differences in the results of simulation
 H1 and T1 in a more compact way by presenting the distribution of gas in
 radial bins. The blue and red dashed curves show the initial ambient gas mass
 within concentric spherical shells of 10 pc width for the turbulent and the
 homogeneous (control) runs, respectively. The solid curves of the same colour
 show how these gas distributions evolve by time $t\simeq 283$ kyr. Note that
 the bubble swept up {\it all} of the ambient gas within a radius of $\sim
 0.45$ kpc into a dense shell in the control run, but is obviously having
 great difficulties in removing the gas in the turbulent simulation. The
 density of the gas in the inner regions actually increases in the latter
 simulation as some of the cold dense gas falls in while the hot bubble
 fizzles out through the pores in the ambient gas.

 These results illustrate clearly the main thesis of our paper: {\it the
   impact of an UFO on the inhomogeneous multiphase medium is much less
   efficient than expected based on spherically symmetric modelling}. 

\section{Dynamics of clumpy gas}

\subsection{Gas dynamics as a function of its density}\label{sec:density}

 We shall now analyse the response of the ambient gas to the presence of the
 hot bubble in the turbulent simulation T1 in greater detail.  This response
 is a strong function of the properties of the ambient gas. Fig.
 \ref{vel-dist} shows the distribution of gas over radial velocity at time
 $t\simeq 70.8$ kyr, for three different initial density ranges
 (i.e. particles are grouped based upon their density at $t=0$). To avoid
 confusion due to dense gas phase shielding lower density gas behind it  and
 therefore unaffected by the feedback flow yet, we include only the SPH
 particles that were within $0.3\leq R\leq 0.35$ kpc at $t=0$ kyr. The red
 and blue histograms show particles that originally have the highest and lowest
 densities whilst the grey curve shows particles at the logarithmic mean
 density. Each of the histograms accounts for $\sim 10\%$ of the total number
 of particles within $0.3\leq R\leq 0.35$ kpc at $t=0$
 kyr. Fig. \ref{vel-dist} demonstrates that the lowest density gas is
 accelerated to high radial velocities, with a mean of $\left<v_{\rm
   r}\right>\simeq 661$ km s$^{-1}$. In contrast, the highest density gas is,
 on average, continuing to infall, with a mean $\left<v_{\rm r}\right>\simeq
 -145$ km s$^{-1}$. The logarithmic mean density gas shows a variety of
 behaviours from an infall with velocity of a few hundred km s$^{-1}$ to an
 outflow with a similar range in velocities.

Also plotted are lines indicating the mean radial velocity of all of the gas
originally in the $0.3\leq R\leq 0.35$ kpc region in the turbulent simulation
($\left<v_{\rm r}\right>\simeq 125$ kms$^{-1}$) and in the homogeneous control
run. In the later case the gas is accelerated to high velocities on average
($\left<v_{\rm r}\right>\simeq 563$ kms$^{-1}$), in a single spherical shell
of swept up material whilst in the turbulent simulation the hot bubble can
escape through the porous medium and so much of the material does not get
accelerated outwards. For the turbulent simulation, not only does the outflow
fail to clear out the high-density material, a large fraction of the
low-density material is also left behind due to shielding by high-density
material in front of it.

\subsection{The column density perspective}\label{sec:column}

 Whilst Fig. \ref{vel-dist} highlights that gas of different densities is
 affected by the outflow differently, it also shows that there is an overlap
 in their radial velocities: some low-density gas is infalling whilst some
 high-density gas is outflowing. This behaviour may partially be due to gas at
 larger radii being shielded from the feedback by dense gas at smaller
 radii. To remove this self-shielding effect in our analysis somewhat, we
 consider the column density of the gas calculated as the integral
\begin{equation}
\Sigma = \int_0^{R} dr\rho(r, \Theta, \phi)\;,
\label{sigma_def}
\end{equation}
along the lines of sight (defined by the spherical coordinate angles $\Theta$
and $\phi$) from the centre of the galaxy.

\begin{figure}
\vskip - 0.5cm \psfig{file=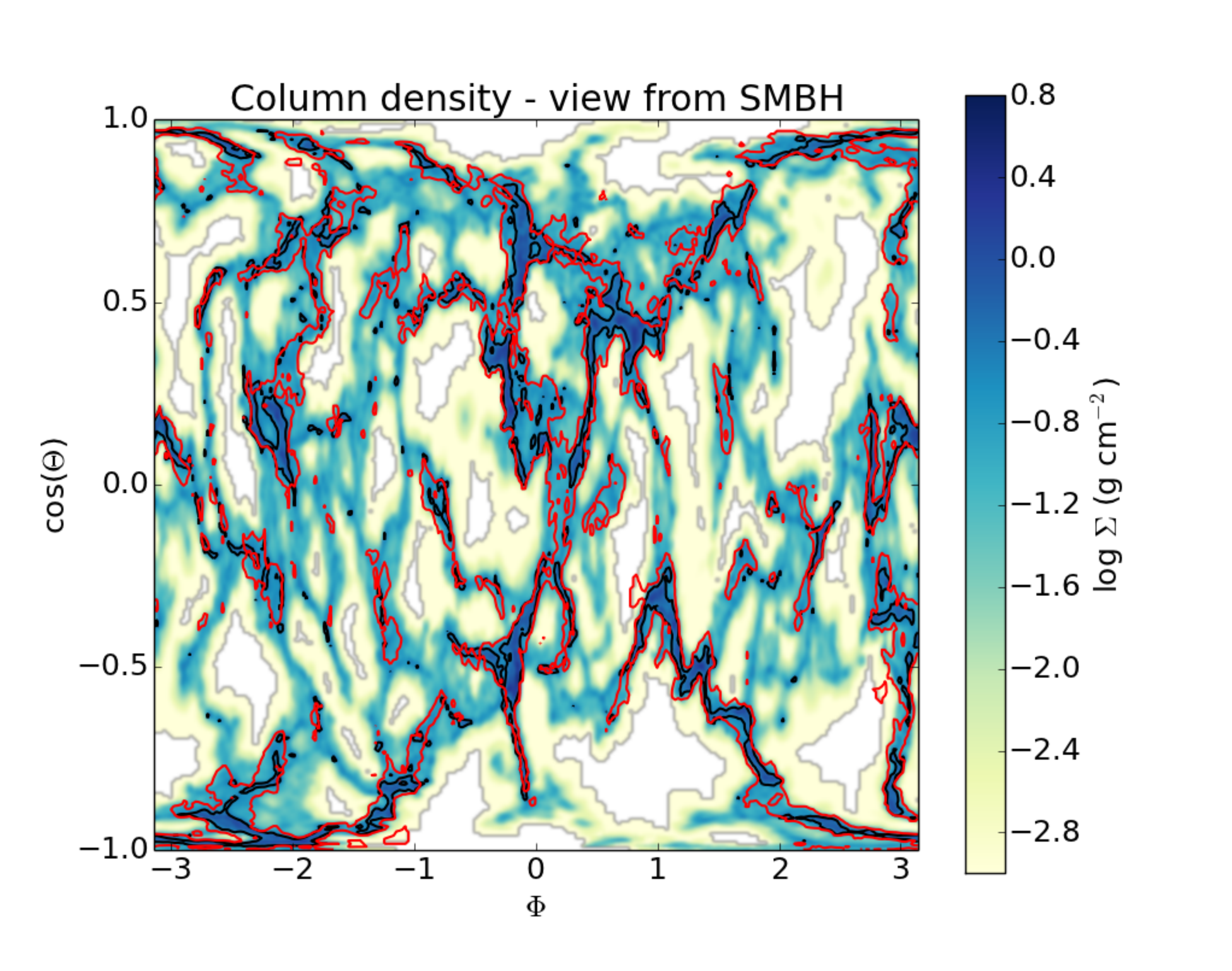,width=0.5\textwidth}
	\caption{Column density of ambient gas at $R\leq 0.35$ kpc at $t=70.8$
          kyr, as viewed from the position of the sink particle. Also plotted
          are contour lines for zero velocity gas (red) and gas with
          $\Sigma_{\rm crit} = 0.36$ g cm$^{-2}$, which is analytically
          predicted to have zero velocity at this time. Note that the two
          contour lines coincide over most of the plot.}
	\label{col-depth}
\end{figure}

 Fig. \ref{col-depth} shows the column density map as a function of the
 position on the sky as viewed from $R=0$. Only ambient gas located inside
 $R\leq 0.35$ kpc at $t=70.8$ kyr is taken into account in this analysis. The
 column density of the ambient gas, $\Sigma$, calculated in this way, varies
 by a factor of about 1000 in Fig. \ref{col-depth}.

Fig. \ref{col-depth} also presents gas radial velocity information by showing
contour lines for zero velocity gas (red). Material inside of these contours
has a negative radial velocity at this time. We can see that it is the gas
with the highest $\Sigma$ that remains infalling, whilst gas with a low
$\Sigma$ generally has positive radial velocities.

The complex nature of gas dynamics in the turbulent simulation makes defining
and analysing the exact dynamics of gas difficult if not impossible since gas
density changes during the simulation. Some of the gas may even switch phases
when it cools or heats up. However we can carry out an approximate analysis by
considering the momentum equation for a clump,
\begin{equation}
\frac{d}{dt}\left(m_{cl}v_{cl}\right)=\pi r_{cl}^{2}P_{\rm ram}
-\frac{Gm_{cl}M(R)}{R^{2}}
\label{mom-eq}
\end{equation}
where $r_{cl}$, $m_{cl}$ and $v_{cl}$ are the clump's radius, mass and radial
velocity, respectively, $P_{\rm ram}$ is the hot bubble's ram pressure acting
on the clump, $R$ is the radial position of the clump and $M(R)$ is the mass
of material within $R$. Making the assumption that $m_{cl}$ and $r_{cl}$
remain approximately constant we can divide through by $m_{cl}$, and re-write
equation \ref{mom-eq} as
\begin{equation}
a_{cl}=\frac{P_{\rm ram}}{\Sigma_{cl}}-a_{grav}
\label{mom-acc}
\end{equation}
where $a_{cl}$ and $a_{grav}$ are the clump's acceleration and gravitational
acceleration, respectively, and $\Sigma_{cl}=m_{cl}/\pi r_{cl}^{2}$ is the
column density of the clump. The ram pressure of the hot gas cannot be predicted
exactly by the analytical model, but we assume that hot gas streams out of its
initial spherical configuration at approximately the sound speed of the hot
gas through numerous ``holes'' in the cold ambient gas distribution. This
argument suggest that by the order of magnitude $P_{\rm ram}$ should be
comparable to the initial isotropic pressure of the hot gas, $P$.

When $\Sigma_{cl}<<P/a_{grav}$, the driving force of the bubble dominates over
gravity and we can neglect the $a_{grav}$ term in equation \ref{mom-acc},
integrating then gives an estimate for a clumps velocity at time $t$ as 
\begin{equation}
v(t)=\frac{P_{\rm ram}}{\Sigma_{cl}}t + v(0).
\label{mom-acc2}
\end{equation}
Setting $v(t)=0$ we can define a critical column density,
\begin{equation}
\Sigma_{crit}(t) = \frac{P_{\rm ram}}{|v(0)|}t
\label{sig-crit}
\end{equation}
such that only material with $\Sigma > \Sigma_{\rm crit}$ should still be
infalling at time $t$, whereas lines of sight with $\Sigma < \Sigma_{\rm
  crit}$ may be launched in an outflow.

Using the mean radial velocity of gas particles at $t=0$ for $v_{0}$, we find
$\Sigma_{\rm crit}\simgt 0.36$ g cm$^{-2}$ at $t\sim 70.8$ kyr. Black contours
in Fig. \ref{col-depth} show the lines of sight where $\Sigma = \Sigma_{\rm
  crit}$. We see that there is a close agreement between the red (zero
velocity contours) and the black contour lines, suggesting that the
approximate analysis based on equation \ref{mom-acc} does have a certain merit
to it. This could be expected from theoretical studies of how a single dense
gas cloud is affected by a hot bubble \citep[e.g.,][]{McKeeCowie75}, N14. The
column density of the cloud, $\Sigma$, is roughly the product of the mean cloud
density, $\rho_{\rm cl}$, and the physical size of the cloud, $r_{\rm
  cl}$. Therefore, a dense but physically small (small $r_{\rm cl}$) cloud may
have a smallish $\Sigma$, and is accelerated to a significant radial velocity
by the UFO, and hence may be completely destroyed, despite being dense.  A
dense and large (large $r_{\rm cl}$ and $\Sigma$) cloud, on the other hand,
may both withstand the onslaught from the hot bubble and also continue to
infall.

There are a few caveats to this approach for comparing $\Sigma$ and expected
radial velocity. The high $\Sigma$ regions shown in the plot can only be
considered an estimate for the high-density material as they are calculated
based upon the entire contribution of material along a particular line of site
out to $R\leq 0.35$ kpc. This leads to potentially over(under)estimating
$\Sigma$ if the clump extends to radii that are less (greater) than $0.35$
kpc. Further we use an average value for $v_{0}$ and assume that the column
density of the clump remains approximately constant over the time period
considered. Therefore the estimate here should only be considered as a rough
illustration of the interaction of the high-density clumps with the expanding
bubble and not an exact solution, which would require a far more detailed
analysis than is necessary for the purposes of this paper.

\subsection{Time evolution of the outflow}\label{sec:phases}

\begin{figure}
\vskip - 0.5cm \psfig{file=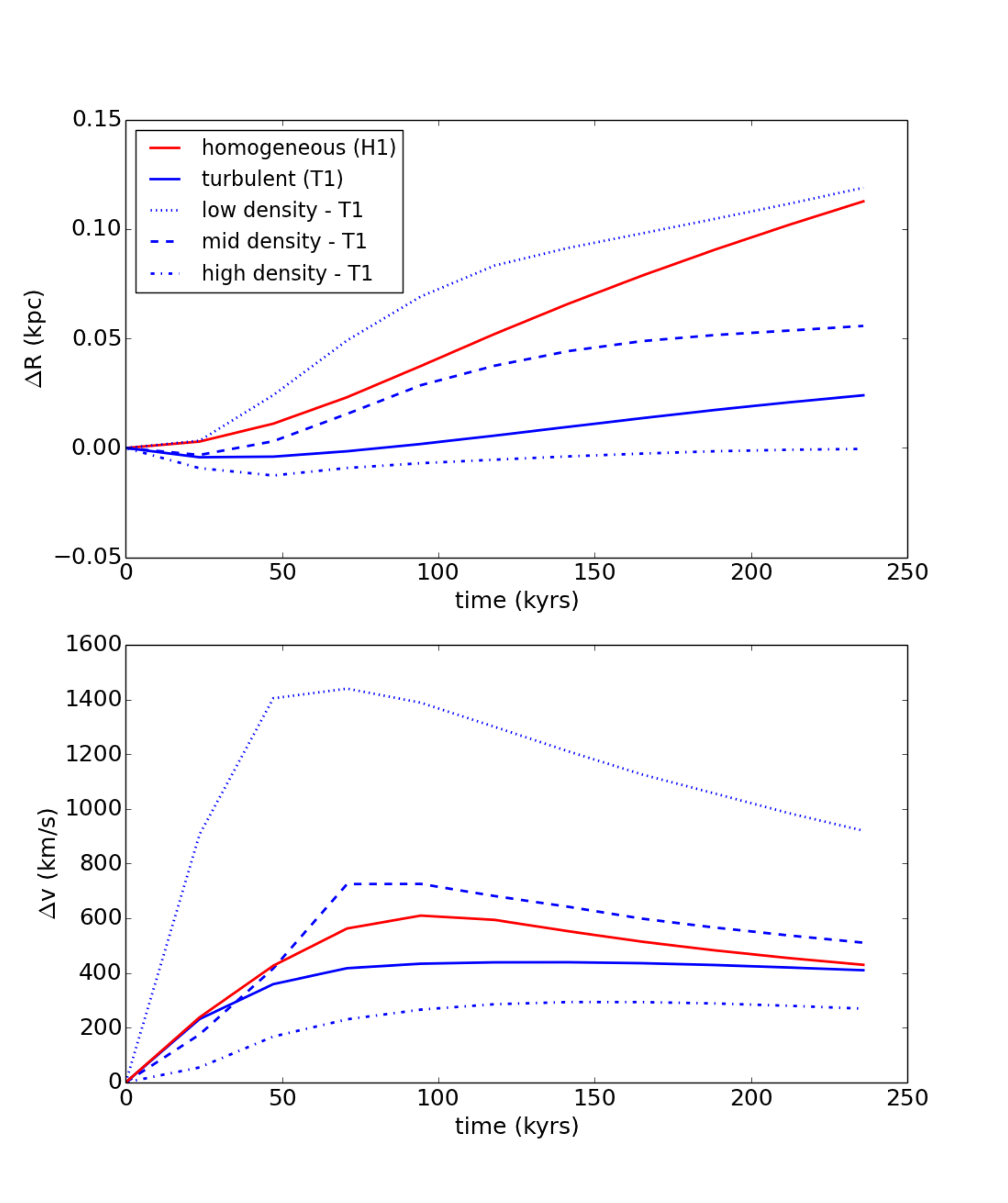,width=0.5\textwidth}
	\caption{Time evolution of the change in mean radial position (top)
          and change in mean radial velocity (bottom) of gas in the
          homogeneous (H1, red) and turbulent (T1, blue) runs. In the latter
          case the gas is further divided into low (dotted), intermediate
          (dashed) and high (dash-dot) density material.}
	\label{btf}
\end{figure}

So far we have only shown properties of the system at specific moments in
time, however, a consideration of the time evolution of the system is also
important. Fig. \ref{btf} shows the time evolution of the change in mean
radial position, $\Delta R=\overline{R}(t)-\overline{R}(0)$ (top) and change in
mean radial velocity $\Delta v = \overline{v}(t) - \overline{v}(0)$ (bottom)
for ambient gas particles initially at $R\le 0.35$ (these particles are chosen
to avoid other complicating factors such as shielding of low density gas). The
solid red and blue lines on these figures are taken from the homogenous
simulation H1 and turbulent simulation T1, respectively. Also shown on each of
the panels in Fig. \ref{btf} is three blue lines calculated from the data of
the turbulent simulation T1, showing the change in mean radial position (top)
and mean radial velocity (bottom) for low (dotted), intermediate (dashed) and
high (dot-dashed) density gas. We apply fixed density thresholds set at the
values used in Fig. \ref{vel-dist} earlier, however, unlike in
Fig. \ref{vel-dist}, where particles are grouped based upon their original
density, here the particles are grouped based upon their density at time
$t$. Both the change in mean radial position and mean radial velocity plots
demonstrate again that the low-density gas is affected by the hot bubble much
stronger than the high density gas. Both panels of Fig. \ref{btf} show a
certain reduction in the difference between the three density groups as time
goes on which is however due to (a) mixing between the two phases with time,
and (b) the fact that the bubble energy is not replenished in our simulation. 

\subsection{Decoupling of energy and mass flow}\label{sec:decoupling}

In the homogeneous control simulation H1, both mass and energy are flowing
outward as the bubble expands. The situation is bound to be far more
interesting in the case of the turbulent simulation T1, since we saw in
Section  \ref{sec:density} that there is both an inflow and an outflow at the
same time. Furthermore, since the different phases have widely different
radial velocities and temperatures, the overall direction of the flow of mass
and energy is not obvious.

\begin{figure}
\vskip - 0.5cm \psfig{file=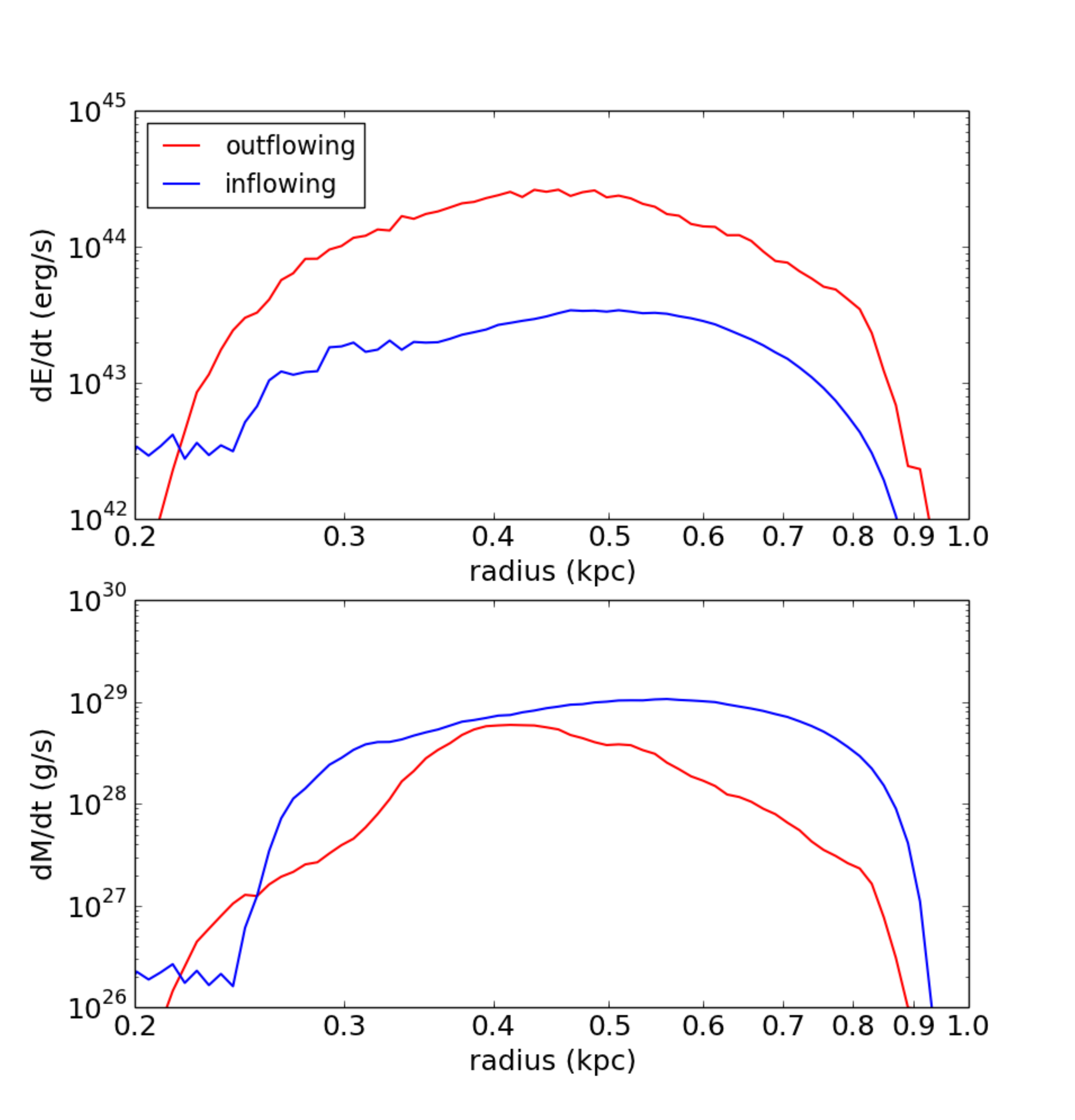,width=0.5\textwidth}
	\caption{Radial flows of energy, $\dot{E}$ (top panel), and mass,
          $\dot(M)$ (bottom panel), for gas that is either in-flowing (blue)
          or outflowing (red) at time $t=283$ kyr in the simulation T1.}
	\label{meflux}
\end{figure}

\begin{figure*}
\vskip - 0.5cm  \psfig{file=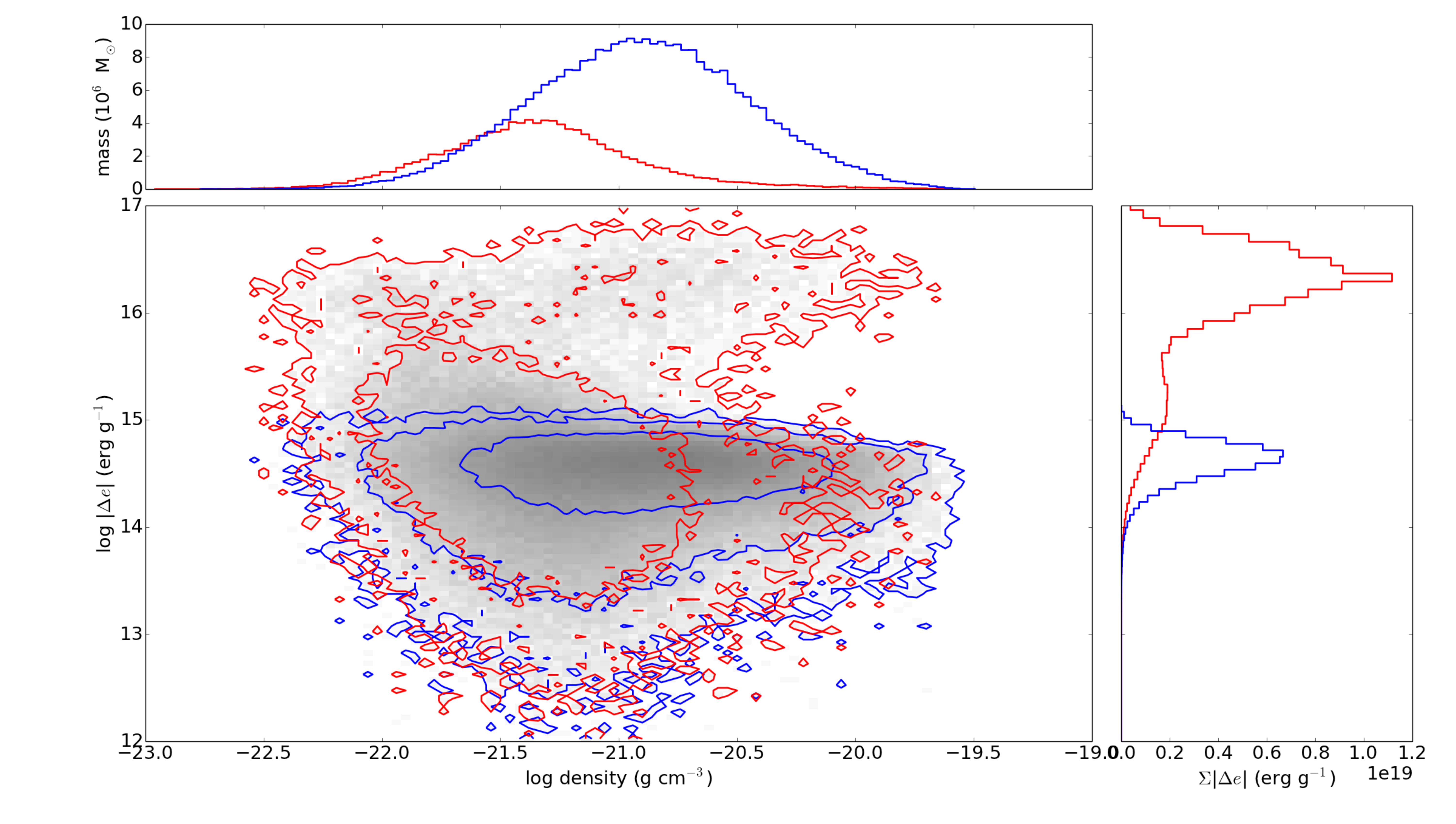,width=0.8\textwidth}
\psfig{file=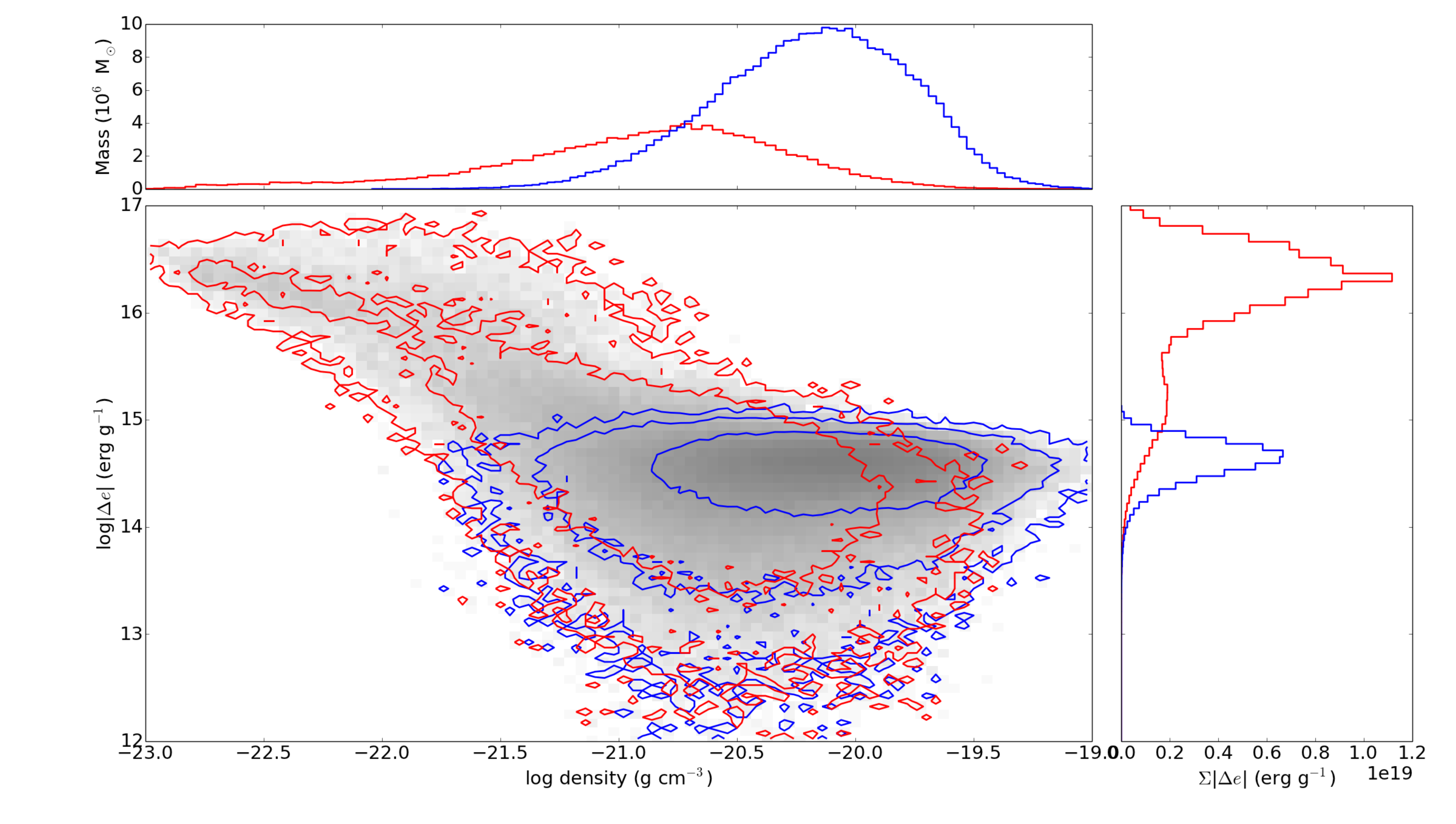,width=0.8\textwidth}
	\caption{Particle distribution plot of absolute change in specific
          energy between $t=0$ and $189$ kyr ($\Delta{e}$) against original
          gas density (top) and current gas density (bottom). Contours
          indicate gas that has lost energy (blue) or gained energy (red). The
          density axis have been collapsed into one-dimensional mass
          histograms above each panel whilst the energy axis has been
          collapsed into one-dimensional histograms weighted by $\Delta{e}$ to
          the right of each panel.}
	\label{dE-rho}
\end{figure*}

To analyse these flows we define the rate of mass and energy flows in a given
radial bin of width $\Delta{r_{bin}}$, respectively, as
\begin{eqnarray}
\dot{M} = \displaystyle\sum{\frac{m_{sph}v_{r}}{\Delta{r_{bin}}}}\\
\dot{E} =
\displaystyle\sum{\left[\frac{1}{2}v^{2}+\frac{3}{2}\frac{k_{B}T}{\mu
      m_{p}}\right]\frac{m_{sph}v_{r}}{\Delta{r_{bin}}}}\;.
\end{eqnarray}
The SPH particles in this sum are selected based on criteria placing them in
one or the other phase or group (see below). In a steady state spherically
symmetric flow, these definitions would include all of the SPH particles in a
bin, and would then give the total mass and energy flux rate as a function of
position in the flow.

 In the homogeneous control run, the energy and mass flows are dominated by
 outflowing material but only within the radius of the swept up shell, beyond
 this there is no outward $\dot{E}$ and $\dot{M}$, while the inward values are
 negligibly small.

Fig. \ref{meflux} shows $\dot{E}$ (top) and $\dot{M}$ (bottom) for in-flowing
($v_{r}\le -\sigma /2$, blue) and outflowing ($v_{r}\ge\sigma/2$, red)
material in the turbulent simulation T1, binned radially at $t=283$ kyr. Both
panels show that, unlike the spherically symmetric situation (simulation H1),
there are outflows and inflows of mass and energy for all radii in the clumpy
simulation T1. Interestingly, the energy flow is dominated by the material
streaming outward, which we identify with the hot low-density gas based on our
earlier analysis, whereas the mass flow is mainly inward and is dominated by
the high density gas.  This shows that {\it energy and mass flows separate
  from one another in turbulent flows.} Unlike the spherically symmetric
homogeneous case, energy does not necessarily flows where most of the mass
does.

To analyse this energy-mass decoupling further, we define the absolute change
in specific energy of SPH particles as
\begin{equation}
|\Delta{e}| =
\left|\frac{1}{2}\left(v^{2}-v_{0}^{2}\right)+\frac{3}{2}\frac{k_{B}}{\mu
  m_{\rm p}}
\left(T-T_{0}\right)+G\frac{M_{a}}{a}\ln\left({\frac{R}{R_{0}}}\right)\right|
\label{e_specific}
\end{equation}
where the terms on the right hand side are the change in specific kinetic,
internal and gravitational potential energy, respectively (note we only
include the gravity due to the underlying potential). $v$, $T$ and $R$ are the
velocity, temperature and radial positions of each particle, respectively,
with the subscript $0$ indicating the initial value of each of these
parameters.

Fig. \ref{dE-rho} shows the absolute change in SPH particle specific energy
($|\Delta e|$) between $t=0$ and $189$ kyr versus the gas density at the
initial time (the top panel), and, alternatively, versus the gas density at
$t=189$ kyr (the bottom panel). Contours indicate gas that has lost energy
(blue) or gained energy (red). The density axis has been collapsed into
one-dimensional mass histograms, located at the top of each plot, whilst the
energy axis has been collapsed into one-dimensional histograms weighted by
$\Delta{e}$, located to the right of each plot. As before (e.g., Fig.
\ref{vel-dist}), only particles within $R=0.35$ kpc at $t=0$ are selected for
this analysis to minimize complications due to gas self-shielding.

Since gas in simulation T1 is initially infalling due to our initial
conditions, so that radial velocity $v_r<0$, particles that loose specific
energy (blue colour in fig. \ref{dE-rho}) correspond to particles that are, in
general, only moderately affected by the hot bubble. The radial velocity of
such particles is either still negative but less so than initially or has a
small positive value.  On the other hand, particles with a positive energy
change (red), as a rule, are particles that are now outflowing with a larger
positive $v_r$.  

Focusing on the 1D mass distributions, above the corresponding panels, we
observe from the figure that most of the mass is in the blue gas that is on
average denser than the red (outflowing) gas. At the same time, 1D energy
distributions to the right of each panel, show that most energy is in the red
SPH particles, so that, consistent with Fig. \ref{meflux}, energy is mainly in
the low-density outflowing particles. The low-density tail of the distribution
of the red particles in the bottom panel shows that the energy gained by the
outflowing particles may be about two orders of magnitude higher than the
energy change of the blue dense particles. 

Further, comparing the top and the bottom panels, we see that the low-density
outflowing gas tail in the bottom panel had on average higher density at time
$t=0$. This gas is initially moderately dense but has been ablated from the
surface of the clouds and launched in the outflow by the hot bubble. The SPH
particles in the blue part of the distribution had their density increased by
a factor of several. The hot bubble thus compresses most of the dense gas by a
factor of at least a few. This is consistent with earlier results of
\citet{NZ12} (see also \citet{SilkNorman09}) showing that AGN outflows may in
fact trigger star formation in dense cold gas by compressing it to very high
densities.

\section{Discussion}

\subsection{Feedback on a homogeneous versus a multiphase ISM}

We have studied the impact of a thermalized UFO launched by a rapidly
accreting SMBH (modelled as a hot bubble) on the ambient gas of the host
galaxy in two contrasting limits. In the first, the ambient gas is initially
homogeneous and spherically symmetric, whereas in the second limit it is
highly inhomogeneous due to an initially imposed turbulent velocity field.  In
broad agreement with previous work (W12, W13, N14 and ZN14), we find marked
differences in the outcome of this interaction. 

We find that the homogeneous spherically symmetric ambient gas is driven
outward by the hot bubble much in the same way as described by the
energy-conserving analytical models of AGN feedback
\citep[e.g.,][]{King03,king05,King10b,ZK12a,FQ12a}. In such models the ambient
gas is only driven away if the feedback is sufficiently strong and the weight
of the medium sufficiently small. In a stark contrast to this, the turbulent
clumpy ISM cannot easily be described in a 1D language. Because of a large
density contrast between the different phases in the ISM, there is
simultaneously inflowing and outflowing gas streaming throughout the host
galaxy. 

The cold dense medium is affected by the UFO significantly less than analytic
models, quoted above, assume because the medium is overtaken by the UFO rather
than being pushed in front of it. We find that some high-density clumps
continue to move inward while the hot bubble fizzles out through low-density
`pores' and accelerates the low-density phase of the ISM to high outward
velocities. Analysis of this behaviour shows that the cold dense phase gets an
initial kick from the pressure of the bubble before it is overtaken, after
which the driving force acting on the clump diminishes. 


Another important result found here is a divergence in the directions of where
most of the mass and energy flow in a turbulent ISM. While most of the mass is
flowing inward, carried by the cold dense clouds which continue to infall
despite AGN feedback, most of the UFO energy manages to percolate through the
ambient ISM and flow outward through the bulge. 

\subsection{Pertinence to the $M-\sigma$ relation}\label{sec:speculation}

Overall, our results suggest that the establishment of the $M-\sigma$ relation
is much more complicated a process than in spherically symmetric models
\citep[e.g.,][]{SilkRees98,Fabian99,King03}. In such models, the $M-\sigma$
mass divides two very different regimes. SMBHs below the $M-\sigma$ mass are
unable to drive the gas outward beyond a small radius (tens to a few hundred
pc, depending on the BH mass and the host velocity dispersion). It is only
once the SMBH exceeds the $M-\sigma$ mass the outflow is able to overcome the
weight of the ambient gas in the galaxy and clear {\it all of the host} of its
gas. This paints an all or nothing picture of AGN feedback (above or below the
$M-\sigma$ mass, respectively).

The picture of AGN feedback changes radically if the ISM in the host is
multiphase. There is no longer the two different regimes with a sharp
boundary, the $M-\sigma$ mass, between them: at any SMBH mass there may be an
inflow and an outflow of gas at the same location in the host and at the same
time.

This must dilute the meaning of the $M-\sigma$ mass, because, on the one hand,
``underweight'' SMBHs, i.e., those below the $M-\sigma$ mass, do have an
influence on the host galaxy even on large scales. Since the hot gas
propagates outward by finding and following the paths of least resistance, the
low-density phase at all radii in the host is vulnerable to AGN feedback. On
the other hand, the high-density medium is more resilient to SMBH feedback
than could be thought based on spherically symmetric models because the medium
is over-taken by the UFO rather than being pushed in front of it. N14 and ZN14
proposed that this unexpected resilience of the host gas to AGN feedback
explains how SMBH manage to grow to the momentum-limited $M_\sigma$ masses
\citep{King03} rather than the energy-limited ($\sim 100$ times lower) masses.

One speculation arising from these results is that a tight $M-\sigma$ relation
could actually never be established in an ensemble of {\it isolated} galaxies,
and that mergers of galaxies are crucial to the emergence of the observed
relations. On the basis of results presented here and in ZN14, we argue that
there are simply too many factors determining the SMBH interaction with the
host galaxy (the ISM structure, angular momentum of the gas, etc.), and that
therefore one should expect a very significant spread in any SMBH-host
relation based on {\it a single episode} of the galaxy and the SMBH growth. It
is likely that averaging occurring during mergers of galaxies \citep[the
  central limit theorem applied to mergers, see e.g.,][]{JahnkeMaccio11}
largely erases this significant spread, leading to a tight $M-\sigma$ relation
at low redshift. This view is consistent with the fact that the observed
SMBH-host scaling relations are only tight for classical bulges and
ellipticals, that the scatter in such relations decreases towards higher
masses, and that SMBH--host relations have larger scatter at large redshifts
\citep{KormendyHo13}.

\subsection{Induced star formation and stellar feedback}

Comparing the top and the bottom panels of Fig. \ref{dE-rho}, or the blue
curves in the horizontal histograms above these panels, we see that the mean
density of the dense material increases with time in the simulation
T1. Because of this density increase, a small number of star particles are
formed in our simulation, in a broad agreement with the earlier suggestions of
AGN-induced star formation \citep{NZ12} (see also \citet{SilkNorman09}). Such
{\it positive} AGN feedback is likely unresolved in cosmological simulations.

\subsection{Comparison with other work}
 Out of previous literature, our work is most similar in spirit to W12 and
 W13, with a number of similar conclusions. One difference, however, is that
 W13 finds that the dense clouds are heated strongly and accelerated outwards
 as a result of the feedback (albeit slower than the hot phase). In our work
 inflows occur {\it despite} the feedback.

The response of the cold phase to the UFO is strongly dependent on the initial
conditions of the phase and the physics included in the simulation.  In W13,
radiative cooling below a temperature of $10^4$~K is turned off, which clearly
limits the highest densities that could be reached by the cold phase under the
external compression by the hot medium. In our simulations, self-gravity of
the clouds is an important factor in ensuring the integrity of the clouds when
they are hit by the UFO. W12 and W13, on the other hand, do not include
self-gravity of the gas and the initial densities of the clouds appear to be
comparable to the tidal densities at the clouds' locations. In our opinion,
cold clouds in W12 and W13 are both susceptible and defenceless to shear from
the gravitational potential and hydrodynamic forces by the UFO.

In any event, we believe that neither our study nor the previous work gives
complete and {\it quantitatively} definitive answers on the interaction of the
UFO and a clumpy turbulent medium of the host galaxy. Future simulations
should focus on modelling the physical properties of the ISM with a greater
realism, in particular including star formation and its feedback (which we did
not include here).

\subsection{Implications for cosmological simulations}\label{sec:cosm}

Cosmological simulations \citep[e.g.,][]{DiMatteo08, Schaye10, Dubois12} often
invoke AGN feedback in order to reproduce observed relationships such as the
galaxy luminosity function. In this sense AGN provide a source of negative
feedback and therefore the mechanism of the sub-grid prescription employed
acts to inhibit star formation and eject gas from a galaxy. This is normally
achieved through heating or ``kicking'' gas local to the black hole. Such
simulations, which by necessity, balance on the edge of what is numerically
achievable, are unable to resolve the multiphase ISM. It is likely that any
feedback would be acting on a single phase medium. The heterogeneous effects
that feedback has on the different phases of a multiphase ISM illustrated in
our simulations may then be lost due to numerical limitations.

The extent to which this poses a problem depends upon the exact nature of the
multiphase ISM \citep{wagner12} and upon the problem that one wishes to
investigate with the cosmological simulations. With regards to meeting
large-scale observational trends, such as the galaxy luminosity function or
$M-\sigma$ relation, the subgrid models employed by cosmological simulations
may be sufficient. However, as shown in this paper, the exact nature of the
ISM does impact how AGN feedback couples with the ambient gas in a galaxy.  In
our simulations, the cold dense phase is mainly affected by the ram pressure
(momentum) of the UFO, whereas the low-density phase bears the brunt of the
UFO's energy content. In contrast, widely used AGN feedback models
\citep[e.g.,][]{DiMatteo08, Dubois12} tend to neglect the physical
state of the gas and instead focus on the proximity of the gas to the
SMBH. Even though cosmological simulations are currently unable to resolve the
ISM, there may still exist material with a range of physical properties close
to black hole. It is therefore likely that the robustness of cosmological
simulations could be improved by a set of prescriptions that incorporate the
physics highlighted by our simulations. Similarly, semi-analytical models
\citep[e.g.,][]{BowerEtal06} may benefit from including an energy-leaking
prescription for the hot bubble (see ZN14).

\section{Conclusion}

We have studied the impact of a thermalized UFO (modelled as a hot bubble)
launched by a rapidly accreting SMBH on the ambient gas of the host galaxy in
two contrasting limits. In the first, the ambient gas is initially homogeneous
and spherically symmetric, whereas in the second limit it is highly
inhomogeneous due to an initially imposed turbulent velocity field.  In a
broad agreement with previous work (W12, W13, N14 and ZN14), we find marked
differences in the outcome of this interaction. In particular, most of the
UFO's energy escapes via low-density channels in the clumpy ISM, which
drastically reduces the impact of the UFO on the dense cold phase that
contains most of the ambient gas in the host galaxy. We conclude that the
state of the ISM in a galaxy is just as important as the AGN feedback model
invoked, in determining how AGN feedback interacts with the ambient medium.

Given the complexity of these processes, the meaning of the $M-\sigma$ mass
becomes much less defined than in spherically symmetric analytic models
\citep[e.g.,][]{SilkRees98,Fabian99,King03}. In the latter, SMBH below the
$M-\sigma$ mass are unable to `clear' their host galaxies and hence continue
to grow, whereas SMBH above this mass terminate their and their host's growth by
expelling all the gas. In a turbulent ISM, there may be outflows -- of the low
density phase -- at $\mbh \ll M_\sigma$, but there could also be inflows -- of
the high-density phase -- at $\mbh \gg M_\sigma$. We therefore concluded in \S
\ref{sec:speculation} that it is hard to see how tight SMBH-host correlations
could occur in an ensemble of {\it isolated} galaxies, and that mergers of
galaxies must be crucial to the emergence of the observed relations. The
interesting question arising from this, then, is to what extent can the
observed correlations be attributed to AGN feedback physics and to what extent
be due to the central limit theorem \citep{JahnkeMaccio11}.

\section*{Acknowledgements}
We acknowledge an STFC grant and an STFC research studentship support. We
thank Hossam Aly, Alex Dunhill and Kastytis Zubovas for useful discussions,
and Justin Read for the use of SPHS. This research used the ALICE High
Performance Computing Facility at the University of Leicester and the DiRAC
Complexity system, operated by the University of Leicester IT Services, which
forms part of the STFC DiRAC HPC Facility (www.dirac.ac.uk). This equipment is
funded by BIS National E-Infrastructure capital grant ST/K000373/1 and  STFC
DiRAC Operations grant ST/K0003259/1. DiRAC is part of the UK National
E-Infrastructure. Figs. \ref{time-evo-uniform} and \ref{time-evo-rho} were
produced using SPLASH \citep{Price07}. 

\bibliographystyle{mnras} \bibliography{nayakshin}
\label{lastpage}

\end{document}